\documentclass[prd,nofootinbib,preprint,superscriptaddress,twocolumn,10pt]{revtex4}
\pdfoutput=1
\usepackage[T1]{fontenc}
\usepackage{amsmath,amssymb}
\usepackage{braket}
\usepackage{epsfig}
\usepackage{graphicx}
\usepackage[usenames,dvipsnames]{color}
\usepackage{subfigure}
\usepackage{slashed}
\usepackage[colorlinks,citecolor=blue]{hyperref}
\usepackage{pdfpages}
\usepackage{color}
\usepackage{comment}
\usepackage{tikz-feynman}
\begin{document}


\title{Electromagnetic leptogenesis with light-heavy sterile neutrinos}

\author{Debasish Borah}
\email{dborah@iitg.ac.in}
\affiliation{Department of Physics, Indian Institute of Technology Guwahati, Assam 781039, India}
\affiliation{Pittsburgh Particle Physics, Astrophysics, and Cosmology Center, Department of Physics and Astronomy, University of Pittsburgh, Pittsburgh, PA 15260, USA}

\author{Arnab Dasgupta}
\email{arnabdasgupta@pitt.edu}
\affiliation{Pittsburgh Particle Physics, Astrophysics, and Cosmology Center, Department of Physics and Astronomy, University of Pittsburgh, Pittsburgh, PA 15260, USA}

\begin{abstract}
We propose a novel leptogenesis scenario utilising the two-body decay of heavy right handed neutrino (RHN) via the electromagnetic dipole operator. While the requirement of the standard model (SM) gauge invariance requires such dipole operator only at dimension-6 forcing the generation of non-zero CP asymmetry from three-body decay with two-loop corrections, we write down dimension-5 dipole operators involving heavy RHN $N_R$ and its lighter counterpart $\nu_R$. This allows the generation of lepton asymmetry in $\nu_R$ from two-body decay of heavy RHN which later gets transferred to left handed leptons via sizeable Yukawa coupling with a neutrinophilic Higgs doublet. The asymmetry in left handed leptons is then converted to baryon asymmetry via electroweak sphalerons. The lepton number violation by heavy RHN also induces a one-loop Majorana mass of $\nu_R$ rendering the light neutrinos to be Majorana fermions. While smallness of the Majorana mass of $\nu_R$ prevents additional sources or washout of lepton asymmetry, it also constrains the scale of leptogenesis. Sub-GeV sterile neutrinos, depending upon their masses come with interesting implications for low energy experiments, neutrino oscillation, warm dark matter as well as effective relativistic degrees of freedom. Additionally, heavy RHN can lead to observable monochromatic photon signatures at terrestrial experiments.
\end{abstract}
\maketitle
\section{Introduction}
The observed baryon asymmetry of the Universe (BAU) has been a longstanding puzzle in particle physics and cosmology \cite{Zyla:2020zbs, Aghanim:2018eyx}. While the standard model (SM) of particle physics fails to address this issue, baryogenesis \cite{Weinberg:1979bt, Kolb:1979qa} as well as leptogenesis \cite{Fukugita:1986hr} have been the most widely studied beyond standard model (BSM) frameworks proposed to explain the observed BAU. In leptogenesis, a non-zero lepton asymmetry is generated first and then gets converted into baryon asymmetry by electroweak sphalerons \cite{Kuzmin:1985mm}. One appealing feature of leptogenesis is its natural realisation within canonical seesaw mechanisms like type-I \cite{Minkowski:1977sc, GellMann:1980vs, Mohapatra:1979ia, Schechter:1980gr, Schechter:1981cv}, type-II \cite{Mohapatra:1980yp, Schechter:1981cv, Wetterich:1981bx, Lazarides:1980nt, Brahmachari:1997cq} and type-III \cite{Foot:1988aq} seesaw proposed to explain non-zero neutrino mass and mixing, another observed phenomena which the SM fails to address. Leptogenesis in such seesaw models typically involve the out-of-equilibrium CP violating decay of a heavy particle into SM leptons via Yukawa portal interactions. Depending upon the specific realisation of leptogenesis, it is also possible to utilise other portals. The authors of \cite{Bell:2008fm, Choudhury:2011gbi} utilised the electromagnetic dipole operator involving a heavy right handed neutrino (RHN) and a SM neutrino to generate lepton asymmetry. However, the requirement of SM gauge invariance at the scale of leptogenesis forces such dipole operators to arise only at dimension-6 level or higher. Consequently, the RHN can generate lepton asymmetry only via three-body decay after two-loop corrections are taken into account.

Motivated by this, we propose a simpler realisation of leptogenesis via the electromagnetic dipole operator by introducing another type of RHN namely, $\nu_R$ which couples to left handed neutrino $\nu_L$ to form a Dirac neutrino at tree level. This allows dipole operator involving $\nu_R$ and heavy RHN $N_R$ at dimension-5 level, responsible for two-body decay of $N_R$ into $\nu_R$ and neutral gauge bosons. Non-zero CP asymmetry is generated after taking one-loop correction to this decay if at least two copies of $N_R$ are considered. The asymmetry generated in $\nu_R$ is then transferred to left handed leptons via Yukawa interactions with the neutrinophilic Higgs doublet \cite{Heeck:2013vha, Borah:2022qln}. Due to the lepton number violation induced by heavy RHN $N_R$, a Majorana mass of $\nu_R$ is also generated at one-loop level. This radiative correction splits the three Dirac neutrinos at tree level into three active and three sterile Majorana neutrinos. Keeping the Majorana mass of $\nu_R$ small to control the washouts of asymmetry created from $N_R$ decay also constrains the scale of leptogenesis and other parameters of the model. The additional light degrees of freedom in neutrino sector, depending upon their masses can have interesting implications for neutrino oscillation, warm dark matter (DM) in the form of keV scale sterile neutrino or enhanced effective degrees of freedom $N_{\rm eff}$ which can be probed at future cosmic microwave background (CMB) experiments. Additionally, one can search for heavy RHN signatures at collider experiments via its decay into monochromatic photon lines due to the dipole operator.

This paper is organised as follows. In section \ref{sec1}, we discuss the basic particle physics framework followed by the details of leptogenesis in section \ref{sec2}. In section \ref{sec3}, we briefly discuss the detection aspects of our model followed by possible UV completion of the minimal framework in section \ref{sec4}. Finally, we conclude in section \ref{sec5}.

\begin{figure*}
    \includegraphics[scale=0.7]{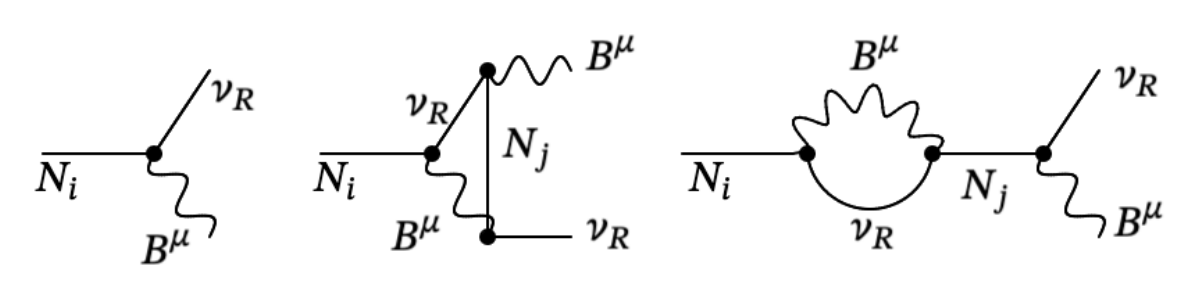}
    \caption{Processes responsible for creating asymmetry in $\nu_R$.}
    \label{fig1}
\end{figure*}

\begin{figure}
    \includegraphics[scale=0.7]{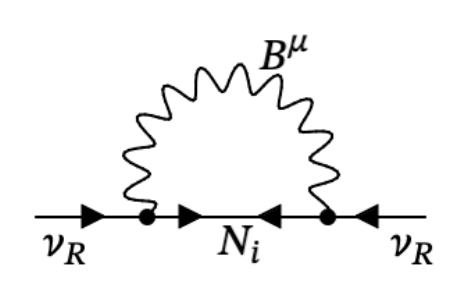}
    \caption{One-loop contribution to Majorana mass of $\nu_R$.}
    \label{fig2}
\end{figure}

\section{The framework}
\label{sec1}
We consider a scenario where the standard model lepton content is extended by two different types of singlet chiral fermions $N_R, \nu_R$. While $\nu_R$ combines with the left chiral neutrinos $\nu_L$ to form Dirac neutrinos at tree level via a neutrinophilic Higgs doublet $H_2$, the heavy right handed neutrino $N_R$ has an effective dipole coupling with $\nu_R$. The relevant Lagrangian is given by 
\begin{align}
   -\mathcal{L} & \supset \frac{\lambda_{i\alpha}}{\Lambda} \overline{N^c_{R_i}} \sigma_{\mu \nu} \nu_{R_{\alpha}} B^{\mu \nu} + \frac{1}{2} M_i  \overline{N^c_{R_i}} N_{R_i} \nonumber \\
   & + y_{\alpha \beta} \overline{L}_\alpha \tilde{H_2} \nu_{R_\beta} + {\rm h.c.} 
   \label{eq1}
\end{align}
where $B^{\mu \nu} = \partial^\mu B^\nu-\partial^\nu B^\mu$ is the field strength tensor of $U(1)_Y$ in the SM. The dipole operator is written as an effective dimension-5 operator suppressed by a cutoff scale $\Lambda$. The validity of the effective field theory (EFT) description requires $\Lambda > M_i$. We assume the heavy RHNs to be in mass basis for simplicity and denote them as $N_i$ without specifying the chirality explicitly hereafter. Fig. \ref{fig1} shows the two-body decay $N_i \rightarrow \nu_R B^\mu$ and possible one-loop corrections responsible for generating the CP asymmetry. The corresponding CP asymmetry can be written as
\begin{align}
    \epsilon_i & = \frac{\Gamma (N_i \rightarrow \nu_R B^\mu) - \Gamma (N_i \rightarrow \overline{\nu_R} B^\mu)}{\Gamma (N_i \rightarrow \nu_R B^\mu) + \Gamma (N_i \rightarrow \overline{\nu_R} B^\mu)} \nonumber \\
    & =\frac{\Gamma (N_i \rightarrow \nu_R B^\mu) - \Gamma (N_i \rightarrow \overline{\nu_R} B^\mu)}{\Gamma_{i}},
\end{align}
where the decay width is given by 
\begin{equation}
    \Gamma (N_i \rightarrow \nu_R B^\mu) =\Gamma (N_i \rightarrow \overline{\nu_R} B^\mu)= \frac{(\lambda^\dagger \lambda)_{ii}}{4\pi} \frac{M^3_1}{\Lambda^2}.
    \label{decaywidth}
\end{equation}
The CP asymmetry generated from the interference of one-loop self-energy and tree level decay process is given by \cite{Bell:2008fm}
\begin{align}
    \epsilon^s_{i\alpha} & = \frac{1}{4\pi} \frac{M^2_i}{\Lambda^2 (\lambda^\dagger \lambda)_{ii}} \sum_{j\neq i}{\rm Im} \bigg ( \lambda^*_{\alpha i}\lambda_{\alpha j} (\lambda^\dagger \lambda)_{ij}\frac{\sqrt{x}}{1-x} \nonumber \\
    & +\lambda^*_{\alpha i}\lambda_{\alpha j} (\lambda^\dagger \lambda)_{ji}\frac{1}{1-x} \bigg),
\end{align}
where $x=M^2_j/M^2_i$ and the second term vanishes after summing over lepton flavor $\alpha$ leading to
\begin{align}
    \epsilon^s_{i} & = \frac{1}{4\pi} \frac{M^2_i}{\Lambda^2 (\lambda^\dagger \lambda)_{ii}} \sum_{j\neq i}{\rm Im} \bigg ( (\lambda^\dagger \lambda)^2_{ij}\frac{\sqrt{x}}{1-x} \bigg).
    \label{esr}
\end{align}
In the resonant limit $M_j-M_i \sim \Gamma_i$, we can enhance the CP asymmetry as 
\begin{equation}
    \epsilon^s_i \sim \frac{\sin{(2\theta)}}{4\pi}
\end{equation}
leading to the resonant leptogenesis \cite{Pilaftsis:2003gt} regime. In deriving the above expression, $\lvert \lambda \rvert$ is assumed to be the common magnitude of the couplings $\lambda_{i\alpha}=\lvert \lambda \rvert e^{i\theta_{i\alpha}}$ with a common relative phase angle given by $\theta=\theta_{i\alpha}-\theta_{j \alpha}$.\footnote{For two generations of heavy RHN $N_i, i=1,2$, the matrix $\lambda_{\alpha i}$ is $3\times 2$ with 6 complex elements. Accordingly, the matrix $\lambda^\dagger \lambda$ appearing in Eq. \eqref{esr} is a $2\times 2$ Hermitian matrix with 1 phase. For three generations of heavy RHN, $\lambda^\dagger \lambda$ is $3\times 3$ Hermitian with three phases in off-diagonal elements.} The CP asymmetry from vertex correction can similarly be evaluated as \cite{Bell:2008fm}
\begin{align}
    \epsilon^v_{i \alpha} & = \frac{1}{4\pi} \frac{M^2_i}{\Lambda^2 (\lambda^\dagger \lambda)_{ii}} \sum_{j\neq i}{\rm Im} [\lambda^*_{\alpha i}\lambda_{\alpha j} (\lambda^\dagger \lambda)_{ij}]\sqrt{x}\bigg [1 \nonumber \\
    & +2x \left (1-(x+1)\ln{\frac{x+1}{x}}\right ) \bigg ].
\end{align}

The asymmetry generated in $\nu_R$ is then transferred to left-handed leptons via sizeable coupling with $H_2$. If this asymmetry is transferred to the left-handed leptons before sphalerons decouple, a non-zero baryon asymmetry can be generated. Neutrinos acquire small Dirac mass at tree level from the vacuum expectation value (VEV) of the neutral component of $H_2$ (denoted by $v_2$) as
\begin{equation}
    M_D = \frac{y v_2}{\sqrt{2}}
\end{equation}
where $v_2$ can be much smaller than the VEV of the SM-like Higgs doublet $H_1$. This can be achieved if the VEV of $H_2$ is protected by an approximate global symmetry and can only be induced after electroweak symmetry breaking due to the VEV of $H_1$. Smallness of $v_2$ compared to $v_1 \equiv \langle H_1 \rangle$ ensures that the Yukawa coupling $y$ is sufficiently large to transfer the asymmetry from $\nu_R$ to $\nu_L$ as we discuss below. 

One can also have a Majorana mass term of $\nu_R$ generated at one loop as shown in Fig. \ref{fig2}. The Majorana mass of $\nu_R$ can be estimated as
\begin{equation}
    M^{\nu_R}_{\alpha \beta} \approx \frac{1}{16\pi^2} \frac{1}{\Lambda^2} \lambda_{i \alpha} M^3_{i} \lambda_{i \beta}. 
\end{equation}
A large Majorana mass of $\nu_R$ can introduce additional source of lepton asymmetry and washout of the asymmetry already produced from $N_i$ decay. As this Majorana mass term is very similar to the decay width of $N_i$ given in Eq. \eqref{decaywidth}, keeping it small also forces $\Gamma_{i}$ to be small, which leads to interesting correlations among the scale of leptogenesis and light sterile neutrino masses.

\begin{figure}
    \includegraphics[scale=0.55]{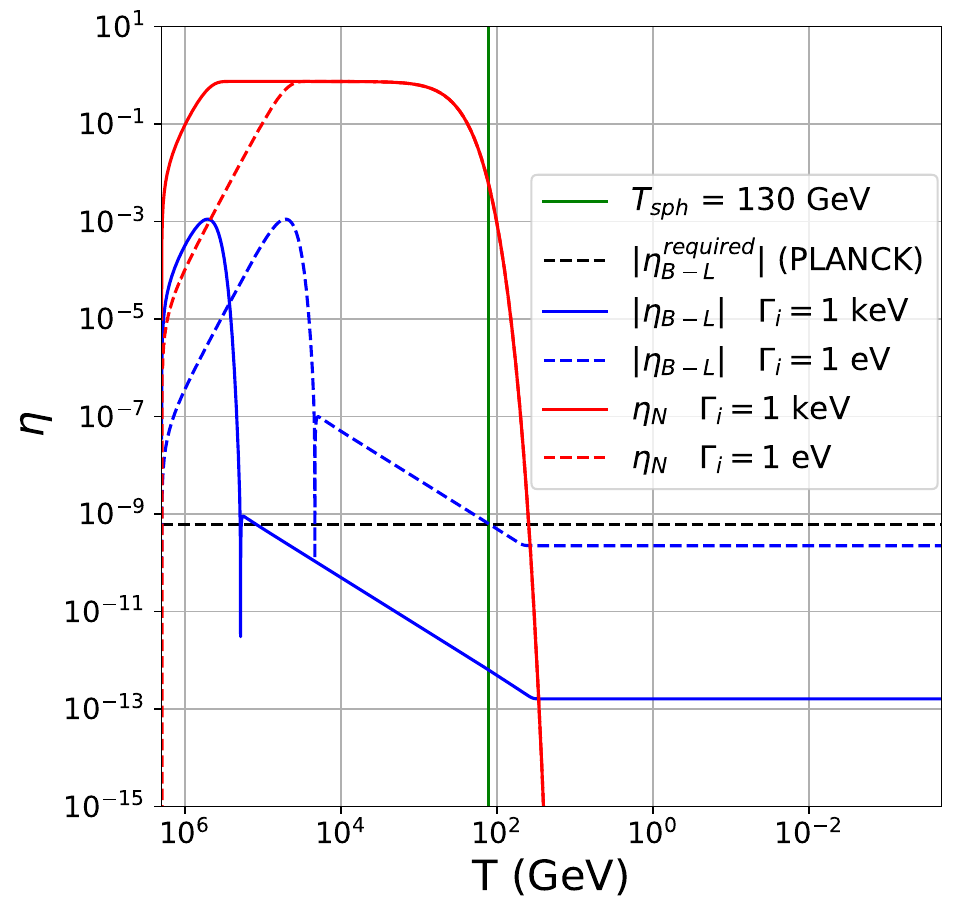}
    \caption{Evolution of comoving abundances of lepton asymmetry and heavy RHN $N_i$ for $\epsilon_1=10^{-2}$, $M_i=1$ TeV, $\Gamma_i=1$ keV (solid), $\Gamma_i=1$ eV (dashed).}
    \label{fig3}
\end{figure}

\begin{figure*}
\begin{tabular}{lr}
    \includegraphics[scale=0.5]{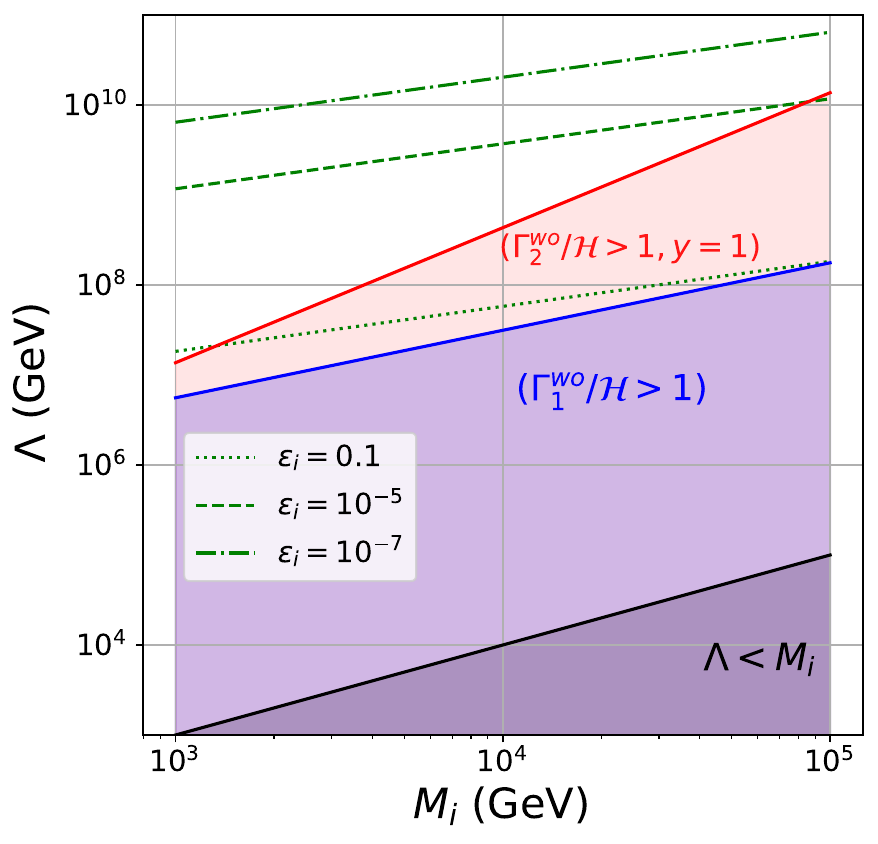} &
    \includegraphics[scale=0.5]{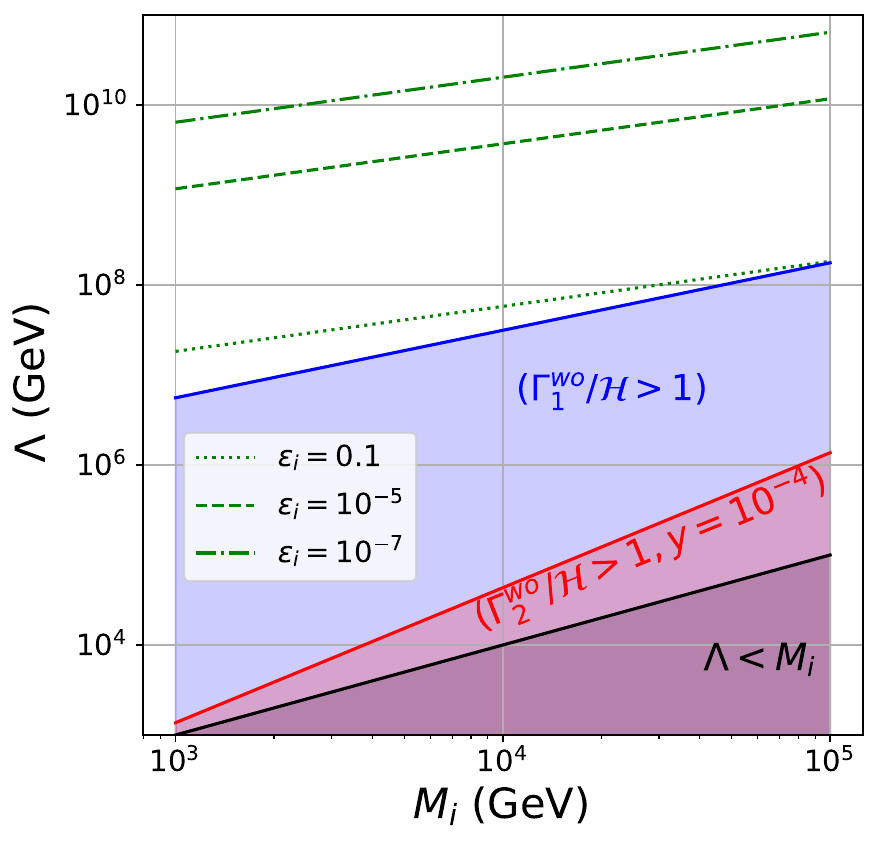}
\end{tabular}
    \caption{Parameter space in $\Lambda-M_i$ plane, considering dimensionless coupling $\lambda \sim \mathcal{O}(1)$. The EFT description is not valid in the grey shaded region $(\Lambda < M_i)$. The blue shaded region is disfavoured due to strong washout from $\nu_R \nu_R \leftrightarrow B^\mu B^\mu$. The red shaded region is disfavoured due to strong washout from $L L \leftrightarrow H_2 H_2$ considering Dirac Yukawa coupling $y=1$ (left panel), $y=10^{-4}$ (right panel).}
    \label{fig3a}
\end{figure*}

\begin{figure*}
\begin{tabular}{lr}
    \includegraphics[scale=0.5]{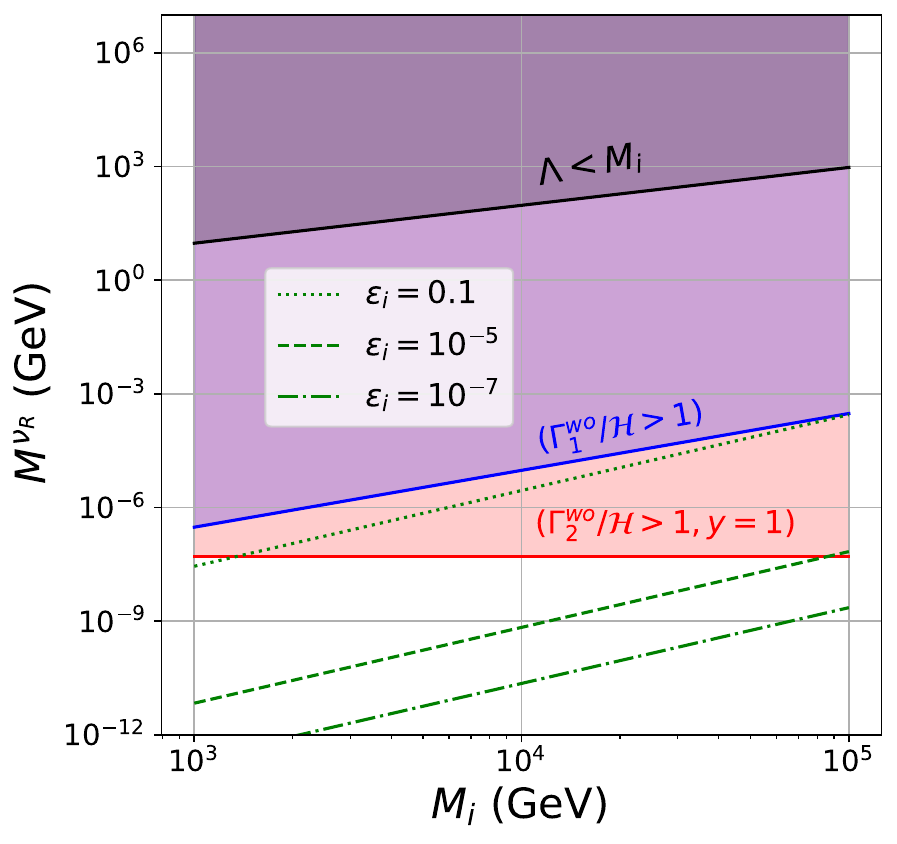} &
    \includegraphics[scale=0.5]{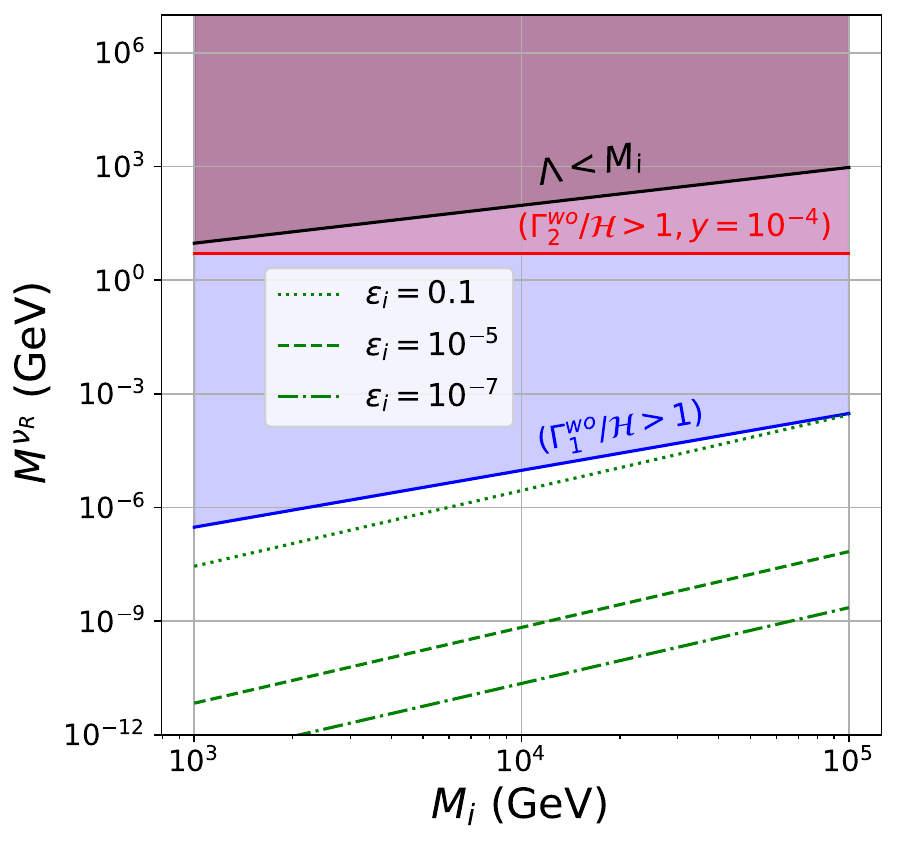}
\end{tabular}
    \caption{Parameter space in $M^{\nu_R}-M_i$ plane, considering dimensionless couplings $\lambda \sim \mathcal{O}(1)$. The EFT description is not valid in the grey shaded region $(\Lambda < M_i)$. The blue shaded region is disfavoured due to strong washout from $\nu_R \nu_R \leftrightarrow B^\mu B^\mu$. The red shaded region is disfavoured due to strong washout from $L L \leftrightarrow H_2 H_2$ considering Dirac Yukawa coupling $y=1$ (left panel), $y=10^{-4}$ (right panel).}
    \label{fig3b}
\end{figure*}

\section{Leptogenesis}
\label{sec2}
Lepton asymmetry sourced from heavy RHN decay via dipole operator, dubbed as Electromagnetic leptogenesis was first proposed in \cite{Bell:2008fm} with follow-up discussions in \cite{Choudhury:2011gbi}. Just with heavy RHN namely, $N_i$, it is not possible to write the gauge invariant magnetic dipole operator at dimension-5 level. This is possible only at dimension-6 level leading to three-body decay of $N_i$ with CP asymmetries arising at two-loop level. Introduction of $\nu_R$ in our setup allows dimension-5 gauge invariant operator keeping the origin of CP asymmetry from two-body decay viable. Discussion of such RHN transition dipole moment as well as CP violation can be found in \cite{Aparici:2009fh, Balaji:2019fxd, Balaji:2020oig}.

The relevant Boltzmann equations can be written as 
\begin{equation}
  \dfrac{d \eta_{\rm N_{1}}}{d z}= -D_{1} \left( \eta_{\rm N_{1}}-\eta_{\rm N_{1}}^{\rm eq} \right),
  \label{eq:BE22}
  \end{equation}
\begin{equation}
 \dfrac{d \eta_{\rm B-L}}{d z}= -\epsilon_{1} D_{1} (\eta_{\rm N_{1}}-\eta_{\rm N_{1}}^{\rm eq})-W \,\eta_{\rm B-L},
\end{equation}
where $\eta_x=n_x/n_\gamma$ is the comoving number density of $x$ and $z=M_1/T, D_1= \langle \Gamma_{1} \rangle/(\mathcal{H}z)$ with $\mathcal{H}$ being the Hubble expansion parameter. The thermal averaged decay width is given by $\langle \Gamma_{1} \rangle = \Gamma_{1} K_1 (z)/K_2 (z)$ where $K_i$'s denote modified Bessel functions of the second kind. $W$ denotes the rates of washout processes including inverse decay and scatterings. The inverse decay rate can be written in terms of the decay rate as $\Gamma_{\rm ID}= \Gamma_{1} (z) \frac{\eta_{\rm N_1}^{\rm eq} (z)}{\eta_{\rm l}^{\rm eq}(z)}$, which gives the washout term due to inverse decay $W_{\rm ID}$ as
\begin{align}
    W_{\rm ID}(z)=\frac{1}{4} K z^3 K_1 (z).
\end{align}
$K$ denotes the decay parameter defined as
\begin{align}
    K=\frac{\Gamma_{1} (z=\infty)}{\mathcal{H} (z=1)}.
\end{align}
The washout term due to inverse decay can be approximately written as \cite{Buchmuller:2004nz}
\begin{align}
    W_{\rm ID}(z)\simeq\frac{1}{4} K z^2 \sqrt{1+\frac{\pi}{2}z}e^{-z}\,\label{eqn:wid2}
\end{align}
which is in equilibrium when $W_{\rm ID}(z)\geq1$. The scattering processes leading to washout include $\nu_R \nu_R \leftrightarrow B^\mu B^\mu$ as one of the dominant ones. The requirement for keeping this washout process out of equilibrium leads to
\begin{equation}
   \Gamma^{\rm wo}_1 \equiv n^{\rm eq} \sigma v < \mathcal{H} \implies T \lesssim \frac{\Lambda^4}{M_{\rm Pl} M^2_1 (\lambda^\dagger \lambda)_{11}}. 
\end{equation}
Adopting a conservative approach of demanding the out of equilibrium of this washout process at $T \leq M_1$ leads to
\begin{equation}
    \Lambda \gtrsim \sqrt{\Gamma_{1} M_{\rm Pl}}
\end{equation}
Since $\Lambda > M_1$ for the validity of the effective dipole operator, the above condition also ensures $K<1$, keeping the washouts under control. Another important washout process is $L L \leftrightarrow H_2 H_2$ or other variants like $L H^\dagger_2 \leftrightarrow \overline{L} H_2$ which violate lepton number by two units. We again take a conservative approach to demand this washout to be out of equilibrium $(\Gamma^{\rm wo}_2 < \mathcal{H})$ for $T \gtrsim T_{\rm sph} \approx 130$ GeV and get an upper bound on the Dirac Yukawa coupling as 
\begin{equation}
    y < \frac{T^{3/4}_{\rm sph}}{M^{1/4}_{\rm Pl} \Gamma^{1/2}_{1}}
\end{equation}
which, for $\Gamma_{1} \sim 1$ keV, leads to $y < \mathcal{O}(1)$. For larger decay width of $N_1$ and hence larger Majorana mass of $\nu_R$, the upper bound will be stronger. The Dirac Yukawa can not be arbitrarily low as the asymmetry in $\nu_R$ gets transferred to the left sector via this Yukawa portal interactions only. Demanding this Yukawa interaction to be in equilibrium prior to the sphaleron decoupling epoch leads to a lower bound 
\begin{equation}
    y \gtrsim 10^{-8}
\end{equation}
which gets stronger if the Yukawa interactions are required to be in equilibrium at higher temperatures. However, as far as the generation of baryon asymmetry is concerned, it is sufficient to ensure that this process enters equilibrium before the sphalerons decouple.

Once the right sector asymmetry is transferred to the left sector, the electroweak sphaleron processes convert the $B-L$ asymmetry into baryon asymmetry with a conversion factor given by 
\begin{align}
   \eta_{\rm B}= \frac{8 N_f + 4 N_\textbf{H}}{22 N_f + 13 N_\textbf{H}}\eta_{\rm B-L} = C_{\rm sph} \eta_{\rm B-L}\,,\label{eqn:sphconv}   
\end{align}
 which for the SM with $N_f=3$ and $N_\textbf{H}=1$, gives $C_{\rm sph} = \frac{28}{79}$. For two Higgs doublets, as in the present setup, $C_{\rm sph} = \frac{8}{23}$. The final baryon asymmetry $\eta_B$ can be analytically estimated to be \cite{Buchmuller:2004nz}
 \begin{align}
     \eta_B = \frac{C_{\rm sph}}{f} \epsilon_1 \kappa\,, \label{eqn:etaana}
 \end{align}
where the factor $f$ accounts for the change in the relativistic degrees of freedom from the scale of leptogenesis until recombination and comes out to be $f=\frac{106.75}{3.91}\simeq27.3$. $\kappa$ is known as the efficiency factor which incorporates the effects of washout processes.

Fig. \ref{fig3} shows the evolution of comoving abundances of heavy RHN and the absolute value of the $B-L$ asymmetry with temperature, assuming the inverse decay as the dominant source of washout. The asymmetry $\eta_{B-L}$ flips sign when RHN abundance increases from $\eta_{N_1} < \eta^{\rm eq}_{N_1}$ to $\eta_{N_1} \geq \eta^{\rm eq}_{N_1}$. This sign flip leads to a dip in the evolution of $\lvert \eta_{B-L} \rvert $ when $\eta_{N_1} \sim \eta^{\rm eq}_{N_1}$. Subsequently, the washouts due to inverse decay lead to a fall in the asymmetry followed by saturation at lower temperatures. Depending upon the decay width $\Gamma_i$ of heavy RHN $N_i$ of mass 1 TeV, the CP asymmetry needs to be tuned in a way such that the required asymmetry at $T=T_{\rm sph}$ survives the washout. Larger decay width leads to larger $K$ and hence more washout requiring enhancement of the CP asymmetry. Since the CP asymmetry is bounded from above $\epsilon \lesssim \mathcal{O}(0.1)$, it also leads to an upper bound on the decay width of $N_i$ for fixed $M_i$. For TeV scale leptogenesis $M_i=1$ TeV, we get an upper bound of around 10 eV for $\Gamma_{i}$ which keeps the Majorana mass of $\nu_R$ at or below $\mathcal{O}(10)$ eV. The comoving abundance of heavy RHN coincides with the equilibrium abundance at an earlier epoch for larger decay width due to enhanced production rates while smaller decay width shows gradual rise in its abundance before reaching equilibrium. The black dashed horizontal line in Fig. \ref{fig3} corresponds to the lepton asymmetry required to create the observed baryon asymmetry quoted by the PLANCK 2018 data \cite{Planck:2018vyg}.

Fig. \ref{fig3a} and Fig. \ref{fig3b} show the allowed parameter space in $\Lambda-M_i$ and $M^{\nu_R}-M_i$ planes respectively by considering the dimensionless coupling associated with the dipole operator $(\lambda)$ to be of order unity. The Dirac Yukawa coupling is taken to be $y=1$ (left panel) and $y=10^{-4}$ (right panel) in these two figures. In the grey shaded region, the EFT description is invalid as $\Lambda < M_i$. The blue shaded region is disfavoured due to strong washout from $\nu_R \nu_R \leftrightarrow B^\mu B^\mu$. The red shaded region is disfavoured due to strong washout from $L L \leftrightarrow H_2 H_2$. Fine-tuning of Dirac Yukawa coupling $y$ weakens the washout constraints related to $\Gamma^{\rm wo}_2$, opening up more allowed parameter space indicated in the right panel plots. From these plots, it is clearly seen that larger CP asymmetry allows smaller (larger) values of cutoff scale $\Lambda$ (Majorana mass of $\nu_R$) compensating for stronger washouts.

\section{Detection Aspects}
\label{sec3}
The leptogenesis framework proposed here has two different types of sterile neutrinos. While the heavy sterile neutrinos at TeV scale or above are responsible for creating the lepton asymmetry, the light sterile neutrinos at sub-GeV scale can have interesting roles in addition to transferring the asymmetries to the left handed lepton sector before sphaleron decoupling. There have been serious attempts to search for such heavy neutral leptons (HNL) with masses above or below the electroweak scale, a recent review of which can be found in \cite{Abdullahi:2022jlv}. While TeV scale HNL in our setup is dominantly produced with a sub-GeV HNL via the dipole operator at hadron or lepton colliders, the latter can also be produced due to mixing with active neutrinos. Heavy sterile neutrino can give rise to mono-photon and light sterile neutrino final state signatures at colliders \cite{Aparici:2009fh}, far facilities like FASER, intensity frontier experiments like SHiP \cite{Barducci:2022gdv} or beam dump experiments like NA62 \cite{Barducci:2024nvd}. On the other hand, sub-GeV HNL can lead to a variety of final states depending upon their masses and mixing with active neutrinos. There are dedicated experiments like NA62\cite{NA62:2020mcv}, SHiP\cite{SHiP:2018xqw}, DUNE \cite{Breitbach:2021gvv}, FASER\cite{FASER:2018eoc}, Codex-b\cite{Aielli:2019ivi}, MATHUSLA\cite{Curtin:2018mvb} to observe such signatures of sub-GeV HNL.

HNL at keV scale, on the other hand, can be a good DM candidate. Depending upon the mixing with active neutrinos or the strength of the dipole operator, such DM can be produced wither via active-sterile mixing \cite{Dodelson:1993je} or conventional freeze-in \cite{Datta:2021elq}. While thermal production of such light DM will typically lead to overproduction, late entropy injection from decay of a long-lived heavier counterpart can bring the relic within observed limits \cite{Bezrukov:2009th, Borah:2021inn}. Such light sterile neutrino DM also falls in the category of warm dark matter where DM can remain semi-relativistic during the epoch of matter-radiation equality. While such DM can have interesting role in alleviating the small-scale structure issues of  cold dark matter, they can also be searched for at X-ray telescopes via monochromatic photon lines. A comprehensive review of such light sterile neutrino DM can be found in \cite{Drewes:2016upu}.

Sterile neutrinos below the keV scale can have several interesting consequences depending upon their mixing with active neutrinos as well as the strength of the dipole operator. Thermalised light sterile neutrinos can give rise to enhancement of the effective relativistic degrees of freedom $N_{\rm eff}$ constrained by CMB experiments like PLANCK as ${\rm N_{eff}= 2.99^{+0.34}_{-0.33}}$ at $2\sigma$ or $95\%$ CL including baryon acoustic oscillation (BAO) data \cite{Planck:2018vyg}. Similar bound also exists from big bang nucleosynthesis (BBN) $2.3 < {\rm N}_{\rm eff} <3.4$ at $95\%$ CL \cite{Cyburt:2015mya}. Both of these cosmological bounds are consistent with the SM predictions ${\rm N^{SM}_{eff}}=3.045$ \cite{Mangano:2005cc}. Future CMB experiment CMB-S4 is expected reach a much better sensitivity of $\Delta {\rm N}_{\rm eff}={\rm N}_{\rm eff}-{\rm N}^{\rm SM}_{\rm eff}
= 0.06$ \cite{Abazajian:2019eic}, taking it closer to the SM prediction. Another future experiment CMB-HD \cite{CMB-HD:2022bsz} can probe $\Delta N_{\rm eff}$ upto $0.014$ at $2\sigma$. In our setup, $\nu_R$ thermalises with the SM leptons at a scale below the scale of leptogenesis via Yukawa interactions with the neutrinophilic Higgs doublet $H_2$. Assuming the decoupling of all three light sterile neutrinos to occur above the electroweak scale, one can obtain $\Delta N_{\rm eff}=0.14$ \cite{Abazajian:2019oqj} within reach of future CMB experiments. Light eV scale sterile neutrinos having sizeable active-sterile mixing can be important for neutrino oscillation, providing a solution to the short baseline anomalies \cite{Diaz:2019fwt} though with some conflict with other limits such as the ones from cosmological observations. Such light sterile neutrinos can also have other interesting implications like for neutrinoless double beta decay, astrophysics etc. as summarised in \cite{deGouvea:2006gz}. While such new dipole operators can also introduce additional contribution to the magnetic moment of active neutrinos \cite{Bell:2008fm} tightly constrained from XENONnT bounds \cite{A:2022acy}, we do not have such additional contribution upto one-loop in our setup, keeping it within limits.

\begin{figure}
    \includegraphics[scale=0.5]{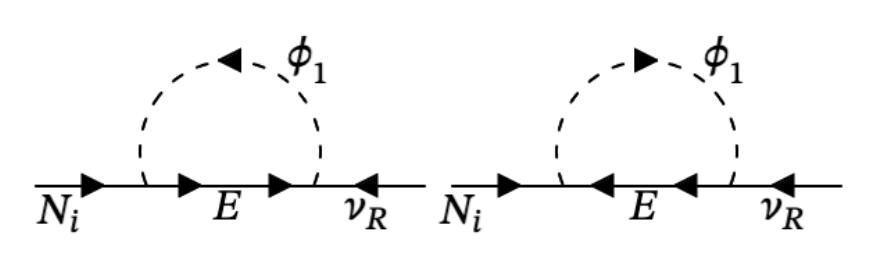}
    \caption{UV completion example 1: Generation of magnetic dipole operator between $N_i$ and $\nu_R$ with charged particles in loop considering $Z_2$ symmetry. The external photon line can originate either from the charged scalar $(\phi_1)$ or charged fermion $E$ inside the loop.}
    \label{fig4}
\end{figure}

\begin{figure}
    \includegraphics[scale=0.5]{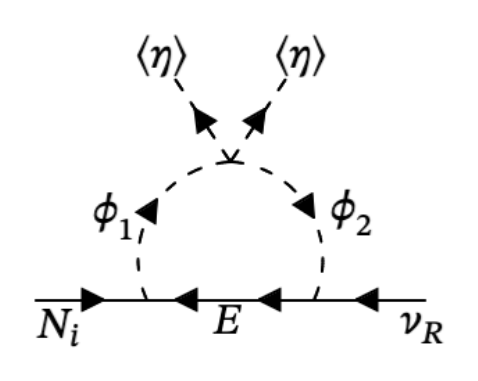}
    \caption{UV completion example 2: Generation of magnetic dipole operator between $N_i$ and $\nu_R$ with charged particles in loop, considering $Z_2 \times Z_5$ symmetry. The external photon line can originate either from the charged scalars or charged fermion inside the loop.}
    \label{fig5}
\end{figure}

\section{UV Completion}
\label{sec4}
The dipole operator given in Eq. \eqref{eq1} can be generated at one-loop level with either scalar-fermion pair or vector boson-fermion pair appearing inside the loop. For example, inclusion of a singlet scalar $\phi_1$ and a vector-like fermion $E_{L,R}$ with hypercharge 1 each leads to the following interactions with $N_R, \nu_R$. 
\begin{align}
    -\mathcal{L}_Y & \supset y_N \overline{E} \phi_1 N_i + y_\nu \overline{E} \phi_1 \nu_R + y'_N \overline{N^c_i} \phi^\dagger_1 E \nonumber \\
    & + y'_\nu \overline{\nu^c_R} \phi^\dagger_1 E +{\rm h.c.}
\end{align}
A $Z_2$ symmetry under which $E, \phi_1$ are odd while all other fields are even, ensures the absence of unwanted terms in the Lagrangian. This leads to the effective dipole coefficient by virtue of the diagrams shown in Fig. \ref{fig4} as \cite{Aparici:2009fh}
\begin{align}
    \frac{\lambda}{\Lambda} \approx \frac{g'}{64\pi^2} \frac{(y_N y'_\nu - y_\nu y'_N)}{M}
\end{align}
with $M \sim M_E \sim M_{\phi_1}$ being the loop particle's mass.

We can also enlarge the particle content and the symmetries to ensure the absence of other unwanted terms in the Lagrangian given by Eq. \eqref{eq1}. For example, an additional $Z_5$ symmetry can be introduced under which the above-mentioned fields are charged as: $\phi_1 (\omega^3), E (1), \nu_R (\omega^4), H_2 (\omega^4), N_R (\omega^3)$. Two more scalars $\eta, \phi_2$ with $Z_5$ charge $\omega$ each are required to generate the dipole operator at one-loop, as shown in Fig. \ref{fig5}. The particles inside the loop are odd under a $Z_2$ symmetry, similar to the example above. The relevant terms in the Lagrangian can be written as
\begin{align}
    -\mathcal{L} \supset y_\nu \overline{E} \phi_2 \nu_R + y_N \overline{N^c_i} \phi^\dagger_1 E + \lambda_{\phi \eta} \phi^{\dagger}_1 \phi_2 \eta^2  +{\rm h.c.}
\end{align}
The neutrinophilic Higgs doublet $H_2$ is assigned $Z_5$ charge $\omega^4$ in order to allow the Yukawa coupling $y \overline{L}\tilde{H_2} \nu_R$. Separate, non-zero $Z_5$ charges of $N_i$ and $\nu_R$ also ensure that neutrinos in the SM only couple to $\nu_R$ via $H_2$ preventing Dirac Yukawa couplings involving $N_i$ and the SM-like Higgs doublet $H_1$. The $Z_5$ symmetry can be broken by non-zero VEV of $\eta$ in addition to the soft breaking in the two Higgs doublet sector via $\mu^2_{\rm soft} H^\dagger_2 H_1$ term leading to induced VEV of $H_2$. Due to the smallness of soft breaking term $\mu_{\rm soft}$, one can generate a much smaller VEV of $H_2$ compared to the electroweak scale ensuring sizeable Dirac Yukawa coupling necessary for our leptogenesis scenario discussed above. It should be noted that in the limit of unbroken $Z_2$ symmetry, we have a stable charged particle beyond the standard model. For cosmological consistencies, it can be made to decay via small $Z_2$-breaking terms like $\phi_1 \epsilon_{\alpha \beta}\overline{L^c}_\alpha L_\beta$ allowing the charged scalar to decay into SM leptons. Another option is to add an electromagnetically neutral particle, say a singlet scalar $\zeta$ odd under $Z_2$ symmetry such that charged $Z_2$-odd particles can decay into it due to interactions like $\overline{E} \zeta l_R$. This does not require $Z_2$-breaking and the stable neutral singlet scalar $\zeta$ can be cosmologically safe as long as its relic does not exceed the observed dark matter relic.

Instead of discrete global symmetries, UV completions can be based on gauge symmetries as well. The simplest possibility of this type is to consider Abelian gauge symmetries like $U(1)_{B-L}, U(1)_{L_\alpha-L_\beta}$ under which the heavy RHN is also charged. While the particle content may have to be enlarged from the requirement of anomaly cancellations, it can enhance the detection aspects. For example, heavy RHN production at colliders, in such a scenario, takes place dominantly via these extra gauge bosons without any suppression of $1/\Lambda^2$. The decay of such RHN, however, continue to take place via the dipole operator keeping the final states unchanged.

\section{Conclusion}
\label{sec5}
We have proposed a novel leptogenesis scenario by utilising a dimension-5 magnetic dipole operator involving two different types of sterile neutrinos: one heavy ($N_i$) and one light $(\nu_R)$. While the heavy $N_i$'s decay into $\nu_R$ and neutral gauge boson to generate lepton asymmetry in $\nu_R$, the latter transfers the asymmetry into the left sector via sizeable Yukawa coupling with a neutrinophilic Higgs doublet $H_2$. The small induced VEV of $H_2$ also leads to a sub-eV Dirac mass of neutrinos. A Majorana mass term of $\nu_R$ also arises at one loop due to the same dipole operator. Requiring this Majorana mass to be in sub-GeV scale to prevent any new source of lepton asymmetry or washout of the asymmetry produced from $N_i$ decay via dipole operator leads to light physical states of sterile neutrinos having interesting implications at different experiments operating at different frontiers. After deriving the necessary CP asymmetries and solving the Boltzmann equations, we show that the correct lepton asymmetry required for the observed baryon asymmetry can be produced at the TeV scale with resonant enhancement. The presence of the dipole operator at dimension-5 level also allows gauge boson mediated production channel of TeV scale heavy right handed neutrinos at collider experiments. As such heavy right handed neutrinos decay only only to a photon and a light sterile neutrino final states, it is expected to give detectable signatures at present as well as future collider experiments. The minimal scenario we propose here can also be accommodated within UV complete setups, as outlined with a realistic example. Such model building attempts can also open up other interesting avenues in terms of solving other problems in particle physics or opening up new detection frontiers. We leave such detailed studies to future works.
\acknowledgements
The work of D.B. is supported by the Science and Engineering Research Board (SERB), Government of India grants MTR/2022/000575 and CRG/2022/000603. D.B. also acknowledges the support from the Fulbright-Nehru Academic and Professional Excellence Award 2024-25.


\providecommand{\href}[2]{#2}\begingroup\raggedright\endgroup

\end{document}